\documentclass{interact}
\usepackage{hyperref}
\usepackage{amsmath,amssymb}
\usepackage{color}
\usepackage{physics}
\usepackage{cite}
\usepackage{bm}
\usepackage[normalem]{ulem}

\begin{document}

\articletype{ARTICLE}

\title{High-pressure magnetic transition in iron observed via diamond quantum sensing}

\author{
\name{Kouki Yamamoto\textsuperscript{a}\thanks{CONTACT Kouki Yamamoto. Email: kouki.yamamoto@phys.s.u-tokyo.ac.jp\\This is an Open Access article published in \textit{High Pressure Research} (2026).\\The Version of Record is available at \url{https://doi.org/10.1080/08957959.2026.2661257}.}, Kenshin Uriu\textsuperscript{b}, Ryotaro Suda\textsuperscript{a}, Misaki Sasaki\textsuperscript{b}, Mari Einaga\textsuperscript{b}, Katsuya Shimizu\textsuperscript{b,c}, Kento Sasaki\textsuperscript{a}, Kensuke Kobayashi\textsuperscript{a,d,e}}
\affil{\textsuperscript{a}Department of Physics, The University of Tokyo, Bunkyo-ku, Tokyo 113-0033 Japan;
\textsuperscript{b}KYOKUGEN, Graduate School of Engineering Science, The University of Osaka, Toyonaka, Osaka 560-8531 Japan;
\textsuperscript{c} Geodynamics Research Center, Ehime University, Matsuyama 790-8577, Ehime, Japan;
\textsuperscript{d}Institute for Physics of Intelligence, The University of Tokyo, Bunkyo-ku, Tokyo 113-0033 Japan;
\textsuperscript{e}Trans-scale Quantum Science Institute, The University of Tokyo, Bunkyo-ku, Tokyo 113-0033 Japan}
}

\maketitle

\begin{abstract}
Diamond quantum sensors offer high precision and spatial resolution as magnetic probes, making them promising for a wide range of applications. While diamond anvil cells (DACs) can generate extremely high pressures, techniques for magnetometry under such conditions remain limited. By fabricating an ensemble of NV centers directly on the anvil diamond surface, we enable precise magnetic measurements under high pressure. In this work, we employ this NV ensemble to image the stray magnetic field of iron up to 30 GPa, enabling the observation of the magnetic transition ($\alpha-\varepsilon$ transition) in iron.
\end{abstract}


\section{Introduction}
Pressure profoundly alters the physical, chemical, and electronic properties of materials. It can induce structural phase transitions, superconducting transitions, and even quantum phase transitions. In particular, hydride high-temperature superconductors, such as $\mathrm{H_3S}$ with a superconducting transition temperature $T_c=203\,\mathrm{K}$ at 155 GPa \cite{Drozdov:H3S, Einaga:H3S_structure} and $\mathrm{LaH_{10}}$ with $T_c=250\,\mathrm{K}$ at 170 GPa \cite{Drozdov:LaH10, Somayazalu:LaH10}, have attracted significant attention because they exhibit remarkably high transition temperatures approaching room temperature under ultrahigh pressure. Moreover, the mechanism of superconductivity in these hydrides can be explained within the framework of BCS theory \cite{Eremets:review}, stimulating extensive research efforts from both experimental and theoretical perspectives.

Ultrahigh-pressure conditions exceeding 100\,GPa are achieved using a diamond anvil cell (DAC). Diamond is the hardest known natural material, and by pressing two opposing diamond anvils together, extremely high pressure can be generated in the sample chamber between them. A high-pressure apparatus based on this principle is a DAC. Though DACs can reach pressures up to 400 GPa \cite{Yagi:DAC}, the sample chamber is very small. For example, when generating 50\,GPa, the chamber diameter is on the order of 80\,$\mathrm{\mu m}$, and for 100\,GPa it is only several tens of micrometers. As the pressure increases, the chamber size becomes progressively smaller. Because of this limited sample volume, magnetic measurements are highly constrained, making precise magnetometry under high pressure challenging. Moreover, signals from samples of such small size are intrinsically weak, and in measurements such as SQUID magnetometry, the entire cell must be measured together with the sample, requiring subtraction of the large background signal originating from the cell \cite{Drozdov:H3S}, which makes precise measurements challenging.

In this work, we perform magnetic measurements under high pressure using diamond quantum sensing. Diamond quantum sensing employs nitrogen–vacancy (NV) centers, point defects in the diamond lattice, as quantum sensors \cite{Taylor2008:NV, Degen2008:NV, Maze2008:NV, Balasubramanian2008:NV}. NV centers allow optical readout of their spin states via optically detected magnetic resonance (ODMR) using laser and microwave excitation. By measuring the spin energy levels, NV centers can be used as sensors for magnetic fields \cite{Tetienne:NV_DW, Velez:NV_DW, Tsukamoto:NV_ML_mag}, temperature \cite{Acosta:NV_temperature, Kucsko:NV_temperature, Yamamoto:NV_temperature}, and pressure \cite{Hilberer:Pressure, Doherty:pressure, Mai:NV_111_iron, Suda:ND_Pressure}.

Because high-pressure conditions are realized using a DAC, fabricating NV centers on the culet surface of the diamond anvil places the sensor directly in the high-pressure environment, enabling precise magnetic imaging even under extreme pressure. Several studies have demonstrated magnetic imaging under high pressure using DACs with NV centers on the anvil culet, including measurements of the ferromagnetic–paramagnetic transition in iron at tens of gigapascals \cite{Lesik:NV_high_pressure,Hsieh:NV_high_pressure,Mai:NV_111_iron} and detection of the Meissner effect in the hydride high-temperature superconductor $\mathrm{CeH_9}$ at approximately 140 GPa \cite{Bhattacharyya:CeH9}. The combination of diamond quantum sensors and DACs is emerging as a powerful technique for precise magnetic measurements at high pressure, offering strong potential for advancing our understanding of material properties under extreme conditions.

In this work, we use a DAC with NV centers fabricated on the anvil surface to observe the ferromagnetic–paramagnetic ($\alpha-\varepsilon$) transition in iron at tens of gigapascals \cite{Taylor:Fe_Hysteresis} by imaging the stray magnetic field originating from the sample. As the pressure increases, the stray magnetic field gradually decreases, corresponding to the transition from the ferromagnetic to the paramagnetic state. Upon decompression, the stray field recovers, reflecting the reverse transition from the paramagnetic to the ferromagnetic state. Iron is a prototypical ferromagnet that undergoes a pressure-induced magnetic transition at room temperature, making it an ideal benchmark material for magnetic imaging using a DAC integrated with NV centers.

A key aspect of this study is the use of a (111)-oriented anvil diamond. When NV centers are fabricated on (100)-oriented anvils, the NV signal becomes strongly degraded under high pressure due to strain in the diamond lattice \cite{Goldman:Contrast_theory, Bhattacharyya:CeH9}, and the sensor becomes unusable above approximately 50 GPa \cite{Hilberer:Pressure}. In contrast, NV centers embedded in (111)-oriented diamond maintain their signal contrast even under high pressure, enabling precise measurements\cite{Wang:NV_111_magnetite,Bhattacharyya:CeH9}. Previous studies measuring the magnetic transition of iron using (100)-oriented anvils reported that the signal directly above the sample became too weak to analyze\cite{Lesik:NV_high_pressure}. In this work, the use of a (111)-oriented anvil allows us to analyze the magnetic signal even directly above the iron sample\cite{Mai:NV_111_iron}. This achievement will lead to the development of NV-based quantitative magnetic imaging under ultra-high pressure, for example, in hydride high-temperature superconductors\cite{Bhattacharyya:CeH9}.

\section{Method}
\subsection{NV center}
NV centers are one type of point defect in diamond. An NV center consists of a substitutional nitrogen atom replacing one carbon atom in the diamond lattice and an adjacent lattice vacancy. The axis connecting the nitrogen and the vacancy defines the NV axis. As determined by the diamond crystal structure, the NV axis can take four possible orientations.

An NV center behaves effectively as a single atomic-like system with an electron spin $S=1$, giving rise to three spin sublevels $m=0,\,\pm1$. The $m=\pm1$ states undergo nonradiative intersystem crossing with a certain probability during relaxation from the laser-excited state, resulting in reduced photoluminescence. Consequently, when sweeping the microwave frequency, the photoluminescence decreases at the resonance frequencies corresponding to transitions involving the $m=\pm1$ states. Plotting photoluminescence contrast as a function of microwave frequency yields the ODMR spectrum, which exhibits dips at the transition frequencies associated with the $m=\pm1$ spin sublevels. In this way, the spin sublevels can be optically read out.

In the presence of a magnetic field, the $m=\pm1$ states undergo Zeeman splitting. The splitting width is linearly proportional to the magnetic field strength, allowing the NV center to be used as a magnetic-field sensor by extracting this splitting from the ODMR spectrum. The proportionality constant for the Zeeman shift, known as the gyromagnetic ratio $\gamma$, is $\gamma=28\,\mathrm{MHz/mT}$ for NV centers. Thus, the Zeeman splitting between $m=\pm1$ levels is $2\gamma B$, where $B$ is the magnetic-field component along the NV axis.

In the absence of a magnetic field, the energy splitting between the $m=0$ and $m=\pm1$ states is called the zero-field splitting (ZFS), which has a value of 2870\,MHz for NV centers at room temperature and ambient pressure. The ZFS is determined by the spin–spin dipolar interaction between the two unpaired electrons of the NV center. Consequently, when the lattice spacing of the diamond changes, the ZFS parameter $D$ also shifts, enabling NV centers to be used as sensors for temperature and pressure by measuring this shift.

The pressure dependence of the ZFS is strongly affected by hydrostaticity. For example, under hydrostatic pressure, the proportionality coefficient is $\partial D/\partial P=14.6\,\mathrm{MHz/GPa}$  \cite{Doherty:pressure}, whereas under non-hydrostatic conditions, a value of $\partial D/\partial P=9.6\,\mathrm{MHz/GPa}$ has been reported \cite{Hilberer:Pressure}.

\begin{figure}
    \centering
    \includegraphics[width=\linewidth]{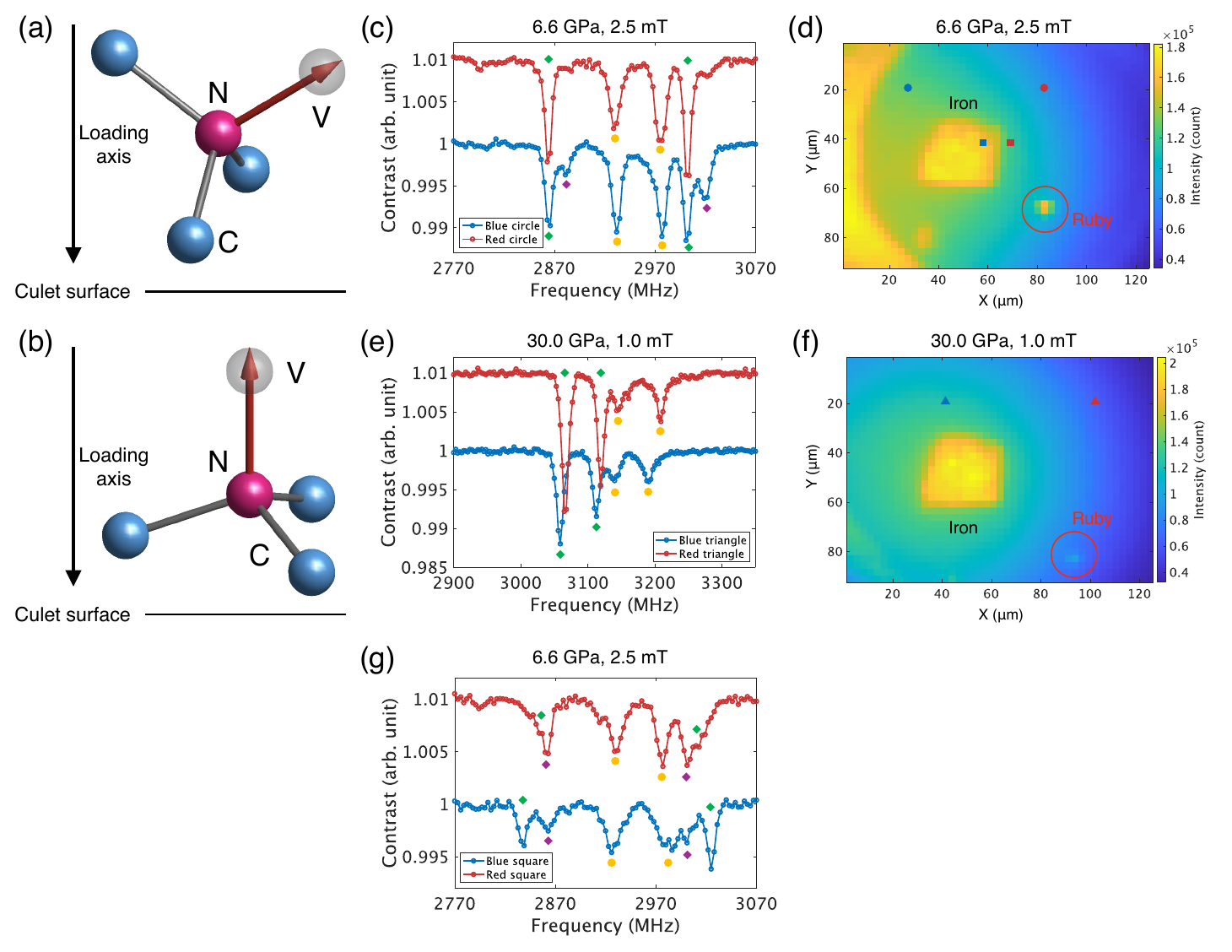}
    \caption{(a) Schematic illustration of an NV center in a (100)-oriented diamond. Pink, blue, and semi-transparent gray spheres represent nitrogen, carbon, and a vacancy, respectively. The red arrow indicates the NV axis. In the (100) orientation, none of the NV axes are perpendicular to the culet surface, making them highly susceptible to strain under pressure. (b) Schematic illustration of an NV center in a (111)-oriented diamond. The NV axis shown is perpendicular to the culet surface, resulting in enhanced robustness against strain. (c) Representative ODMR spectra at 6.6\,GPa with a bias magnetic field of 2.5\,mT. The blue and red spectra correspond to the positions marked by blue and red circles in (d), respectively. Panel (d) shows a bright-field image of the sample chamber (also see Fig. \ref{fig2}(c)). In (c), green diamond markers indicate dips originating from the [111]-oriented NV centers, yellow circles indicate dips from the other three NV orientations, and purple diamond markers indicate dips originating from [111]-oriented NV centers located on the gasket. At positions close to the gasket (blue), clear dips from gasket-originated [111] NV centers (purple diamonds) are observed, whereas these dips disappear at positions farther from the gasket (red). In addition, comparison between the signal originating from the [111]-oriented NV centers (green diamonds) and those from the other three NV orientations (yellow circles) shows that the ZFS associated with the [111] orientation is shifted toward lower frequencies. (e) Representative ODMR spectra at 30.0\,GPa with a bias magnetic field of 1.0\,mT. The blue and red spectra correspond to the positions marked by blue and red triangles in (f), respectively. Panel (f) shows a bright-field image of the sample chamber. In (e), both spectra exhibit four dips; however, the red spectrum is shifted toward higher frequencies. This shift arises from the pressure gradient across the culet surface of the sample chamber. (g) ODMR spectra corresponding to the positions marked by blue and red squares in (d). In the blue spectrum, the stray magnetic field from the iron sample is sufficiently strong that the [111]-derived dip from NV centers on the culet surface (green diamonds) and that from NV centers inside the anvil (purple diamonds) are clearly resolved. In contrast, in the red spectrum, the stray magnetic field from the iron sample is weak, and these dips cannot be resolved. Yellow circles indicate dips originating from the other three NV orientations.}
    \label{fig1}
\end{figure}

\subsection{NV centers fabricated on the diamond anvil culet of a DAC}
To perform magnetic measurements directly in the high-pressure environment, we fabricate NV centers on the culet surface of the diamond anvil. Vacancies are first generated by ion implantation into the culet surface. In this work, we produced vacancies using helium-ion implantation. Subsequent annealing allows the vacancies to diffuse and combine with isolated nitrogen atoms present in the diamond, forming NV centers. In this manner, we fabricated an NV layer on the culet surface with a thickness of about one micrometer.

The crystallographic orientation of the anvil diamond plays a crucial role in NV-based quantum sensing under high pressure. The [100] axis is commonly aligned with the loading axis in high-pressure experiments, but none of the NV axes are perpendicular to the culet surface (Fig. \ref{fig1}(a)), making them highly susceptible to strain under high pressure. Consequently, the NV signal decreases as the pressure is increased, and the sensor becomes unusable above approximately 50\,GPa\cite{Hilberer:Pressure}.

In contrast, for the (111) orientation, one of the NV axes is perpendicular to the culet surface (Fig. \ref{fig1}(b)), and this orientation maintains sufficient signal even under high pressure. Furthermore, because the other three NV orientations exhibit reduced signal under compression, the photoluminescence contrast from the NV axis perpendicular to the culet surface becomes more dominant at higher pressures. Due to these characteristics, (111)-oriented diamond anvils are required for measurements under ultrahigh pressures exceeding 100\,GPa, such as studies of hydride high-temperature superconductors\cite{Bhattacharyya:CeH9}.

In this work, we also employ a (111)-oriented diamond anvil to enable precise magnetic measurements. Owing to polishing constraints, the orientation is not perfectly (111) but is adjusted with an off-angle of several degrees.

\subsection{ODMR spectra under high pressure}
With the NV axis chosen as the $z$-axis along the loading direction, the Hamiltonian $H$ of the NV center under pressure is given by \cite{Hilberer:Pressure,Mai:NV_111_iron,Suda:ND_Pressure,Hsieh:NV_high_pressure}:

\begin{align}
    H &= D_{\mathrm{eff}}\qty(S_z^2-\frac{2}{3}) + E_{\mathrm{eff}}\qty(S_x^2-S_y^2) + E_{\mathrm{eff}}'(S_xS_y+S_yS_x) + \gamma\bm{B}\cdot\bm{S},\\
    D_{\mathrm{eff}} &= D_0 + \kappa P + d_{||}\qty(\sigma_{zz} - \frac{1}{2}(\sigma_{xx}+\sigma_{yy}))\label{Eq_D},\\
    E_{\mathrm{eff}} &= d_{\perp}\qty(\sigma_{xx}-\sigma_{yy}),\\
    E_{\mathrm{eff}}' &= 2d_{\perp}'\sigma_{xy},
\end{align}

where $\gamma$ is the gyromagnetic ratio, $\bm{B}$ is the external field, $\bm{S}=\qty(S_x,\,S_y,\,S_z)$ is the spin-1 operator, $D_0$ is the ZFS at room temperature and ambient pressure, $\kappa$ is the hydrostatic pressure coefficient of $D$, $P$ is the pressure, $d_{||}$ is the coefficient describing the shift of $D$ due to the non-hydrostatic stress component along the NV axis, $d_{\perp}$ is the coefficient describing anisotropic strain (shear stress) in the $x$-$y$ plane, and $d_{\perp}'$ is the coefficient associated with the off-diagonal shear components.

Since the NV centers are fabricated on the culet surface of the anvil diamond, the stress along the [111] direction increases as pressure is applied, leading to a pronounced non-hydrostatic component. The third term on the right-hand side of Eq. (\ref{Eq_D}) represents the effect of non-hydrostatic stress, and because $d_{||}$ is negative\cite{Hsieh:NV_high_pressure}, the effective ZFS $D_{\mathrm{eff}}$ decreases as the stress along the NV axis increases.

Before discussing the experimental details shown in Fig. 2, it is useful at this point to introduce representative ODMR spectra obtained under high pressure. Representative ODMR spectra measured under high pressure are shown below. Figure~\ref{fig1}(c) shows spectra acquired at 6.6\,GPa with a bias magnetic field of 2.5\,mT. The blue and red spectra correspond to the positions indicated by the blue and red circle markers in the bright-field image of the sample chamber shown in Fig.~\ref{fig1}(d), respectively. In Fig.~\ref{fig1}(d), the square object at the center corresponds to the iron sample, and the bright spot at the lower right corresponds to the ruby used for pressure calibration. Also see Fig. \ref{fig2}(c).

In Fig.~\ref{fig1}(c), the green diamond markers indicate dips originating from the [111]-oriented NV centers located above the sample chamber, the yellow circles indicate dips from the other three NV orientations located above the sample chamber, and the purple diamond markers indicate dips originating from the [111]-oriented NV centers located on the culet surface above the gasket. Because the pressure above the gasket is higher than that above the sample chamber, the corresponding ZFS is shifted toward higher frequencies. As a result, at positions close to the gasket, signals originating from NV centers above the gasket (purple diamonds), as shown in the blue spectrum, are superimposed on the spectra.

Furthermore, comparison between the [111]-derived signal above the sample chamber (green diamonds) and the signals from the other three NV orientations above the sample chamber (yellow circles) shows that, under high pressure, the ZFS of the [111]-oriented NV centers is shifted toward lower frequencies.

Next, Fig.~\ref{fig1}(e) shows representative ODMR spectra measured at 30.0\,GPa with a bias magnetic field of 1.0\,mT. The blue and red spectra in Fig.~\ref{fig1}(e) correspond to the positions indicated by the blue and red triangular markers in the bright-field image of the sample chamber shown in Fig.~\ref{fig1}(f), respectively. The green diamond markers indicate dips originating from the [111]-oriented NV centers, while the yellow circles indicate dips from the other three NV orientations. As the pressure is increased, a pressure gradient develops across the culet surface, resulting in position-dependent shifts of the ZFS, as shown in Fig.~\ref{fig1}(e).

Based on these considerations, the dips in the ODMR spectra were successfully assigned to the corresponding NV centers.

\subsection{Influence of NV centers inside the diamond anvil}
In this study, a type-Ib diamond anvil with NV centers fabricated on its culet surface was used. Type-Ib diamond contains a high concentration of impurities, particularly isolated nitrogen atoms, which is advantageous for creating vacancies and subsequently forming NV centers. On the other hand, NV centers may already exist inside the whole anvil prior to ion implantation, and such pre-existing NV centers can act as a background signal and pose a challenge for ODMR spectral analysis.

In the diamond anvil used in this work, a significant number of internal NV centers were present. As a result, in addition to the signal from the NV centers fabricated on the anvil surface to detect the stray magnetic field from the sample, signals from internal NV centers—responding only to the bias magnetic field—were superimposed. When the stray magnetic field from the sample is weak, the ODMR spectra from the surface NV centers and the internal NV centers are not sufficiently separated, making accurate spectral analysis challenging.

Here, we present spectra acquired in a region where a positive stray magnetic field from the iron sample is present in addition to the bias magnetic field (Fig.~\ref{fig1}(g)). The blue and red spectra correspond to the positions indicated by the blue and red square markers in Fig.~\ref{fig1}(d), respectively. While NV centers inside the diamond anvil experience only the bias magnetic field, NV centers on the anvil surface are subjected to the combined magnetic field of the bias field and the stray field from the iron sample.

In the blue spectrum in Fig.~\ref{fig1}(g), the stray magnetic field is sufficiently strong that the signals from the [111]-oriented NV centers inside the anvil (purple diamonds) and those from the [111]-oriented NV centers on the anvil surface (green diamonds) are well separated and can be clearly distinguished. In contrast, when the stray magnetic field from the iron sample becomes weak, as in the red spectrum in Fig.~\ref{fig1}(g), these two signals can no longer be resolved. The yellow circles indicate dips originating from the other three NV orientations.

In this study, spectra exhibiting clear separation between the internal and surface NV signals, such as the blue spectrum in Fig.~\ref{fig1}(g), were fitted using two dips. Conversely, when the two contributions could not be separated, as in the red spectrum in Fig.~\ref{fig1}(g), the spectrum was fitted with a single dip. It should be noted that even when fitting with a single dip in cases where the internal and surface NV signals are not separable, the resulting resonance frequency is shifted relative to the case where only the bias magnetic field is present, due to the contribution from the surface NV centers. Thus, while an accurate estimation of the stray magnetic field magnitude is not possible under these conditions, the presence or absence of a stray magnetic field can still be reliably determined.

To reduce the influence of background signals arising from NV centers inside the diamond anvil, it is important to quantify the density of internal NV centers in advance and to select anvils with a low internal NV concentration before fabricating NV centers on the anvil surface. Alternatively, establishing a method to eliminate or deactivate NV centers inside the anvil is also an important challenge.

\section{Experiment}

\begin{figure}
    \centering
    \includegraphics[width=\linewidth]{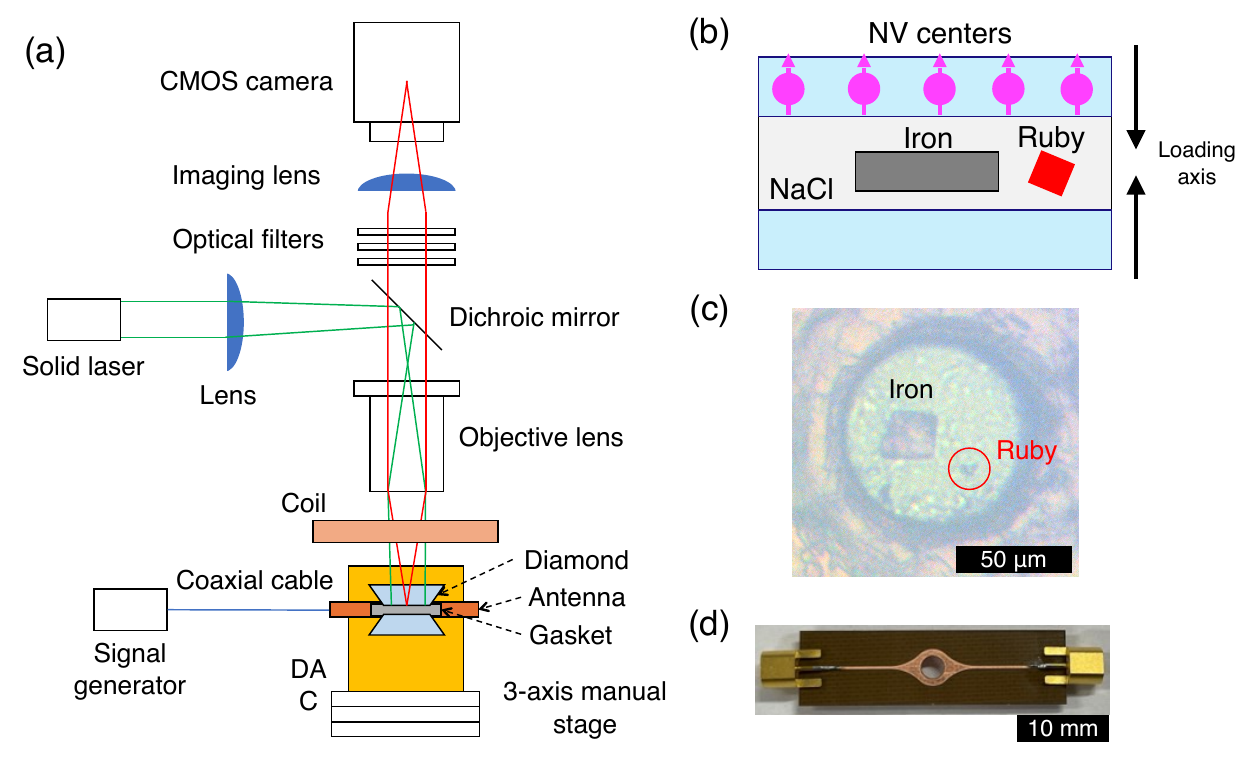}
    \caption{(a) Schematic illustration of the experimental setup. (b) Cross-sectional view of the sample chamber. NV centers are fabricated on the culet surface of the upper anvil diamond. The chamber is filled with NaCl, and both the iron sample and the ruby used for pressure calibration are embedded within the NaCl medium. (c) Optical microscope image of the sample chamber. The square-shaped iron sample is located at the center, and the ruby is enclosed by a red circle. (d) Microwave antenna. Microwaves are applied through the central opening, into which the DAC sample chamber is accommodated.}
    \label{fig2}
\end{figure}

Figure \ref{fig2}(a) shows a schematic of the experimental setup. A 520-nm laser beam was reflected by a dichroic mirror and then focused onto the NV centers on the upper diamond anvil through an objective lens (Mitsutoyo, M-PLAN APO 50$\times$, numerical aperture NA = 0.55, magnification $\times$50). A microwave antenna (Fig. \ref{fig2}(d)) was placed between the two diamond anvils to deliver microwaves for magnetic resonance. The photoluminescence emitted from the NV centers passed back through the objective lens, was transmitted through the dichroic mirror, and was spectrally filtered by a 550-nm long-pass filter and a 600-nm long-pass filter to select the red fluorescence. In the present experiment, a beam splitter was placed upstream of the optical filters to enable spectroscopy of the ruby fluorescence, diverting a portion of the collected light to a spectrometer. The filtered fluorescence was then imaged onto a CMOS camera (Basler, acA720-520um) using a 200-mm focal-length imaging lens.

By acquiring an ODMR spectrum for each pixel of the CMOS camera and analyzing the spectral dips, wide-field magnetic imaging was achieved. During the ODMR measurements, a bias magnetic field was applied to separate the resonance dips. A set of coils installed in the setup enabled the application of a bias field of a few millitesla.

In this study, we measured the spatial distribution of the stray magnetic field associated with the ferromagnetic–paramagnetic transition of iron at room temperature. Iron is in the ferromagnetic body-centered cubic (bcc) phase ($\alpha-$phase) under ambient conditions, but undergoes a structural transition to the paramagnetic hexagonal close-packed (hcp) phase ($\varepsilon-$phase) when compressed to several tens of gigapascals at room temperature. This magnetic transition is commonly referred to as the $\alpha$-$\varepsilon$ transition.

To measure the $\alpha-\varepsilon$ transition, experiments were performed using a DAC equipped with diamond anvils having a culet diameter of 300\,$\mathrm{\mu m}$. A cross-sectional view of the sample chamber is shown in Fig.~\ref{fig2}(b). NV centers were fabricated on the culet surface of the upper diamond anvil. The DAC used in this experiment was an HMD DAC-SPr-30-60 designed for XRD applications.

A nonmagnetic rhenium gasket with an initial thickness of 250\,$\mathrm{\mu m}$ was used. The gasket was preindented to a thickness of 42\,$\mathrm{\mu m}$, after which a sample chamber with a diameter of 80\,$\mathrm{\mu m}$ was drilled using an infrared laser.

NaCl was employed as the pressure-transmitting medium. NaCl was repeatedly loaded into the sample chamber under pressure until it fully filled the chamber and became optically transparent. The iron sample (purity 4N, approximately $35\,\mathrm{\mu m}\times 30\,\mathrm{\mu m}$ in size and $10\,\mathrm{\mu m}$ in thickness) and a ruby sphere for pressure calibration were then loaded and surrounded with additional NaCl to ensure a hydrostatic environment. An optical microscope image of the sample chamber taken through the upper anvil is shown in Fig.~\ref{fig2}(c). The pressure in the chamber was determined using the ruby fluorescence method \cite{Mao:Ruby}.

The microwaves required for the ODMR measurements were applied using the microwave antenna shown in Fig.~\ref{fig2}(d). The central opening of the antenna has a diameter of 2.5\,mm, through which the microwave field is delivered. The culet of the diamond anvil fits into this opening, allowing ODMR measurements of the NV centers on the culet surface.

At each pressure step, an external magnetic field of approximately 37\,mT was applied to the iron sample using a neodymium magnet in order to saturate its magnetization in the desired direction. The iron sample has a plate-like geometry, and its easy axis lies within the plane of the long facet. By magnetizing the sample in this manner, we ensured that any disappearance of the stray magnetic field during compression could not be attributed to laser-induced demagnetization, confirming that the observed change arises from the ferromagnetic–paramagnetic transition. During decompression, a clear reappearance of the stray field was also observed at the point where the sample transitioned back from the paramagnetic to the ferromagnetic state. Note that this external magnetic field was applied only temporarily for magnetizing the sample; during ODMR measurements, only the bias magnetic field was applied.

\section{Result}
\subsection{Magnetization reversal}
We first verified the magnetization reversal of the iron sample by applying an external magnetic field. The magnitude of the applied external field was about 37\,mT, regardless of its direction, which is sufficient to saturate the magnetization of the sample. During the magnetization-reversal measurements, the bias magnetic field was set to 2.5\,mT. The pressure in the sample chamber, determined by the ruby fluorescence method, was 6.6\,GPa, at which iron remains ferromagnetic.

\begin{figure}
    \centering
    \includegraphics[width=\linewidth]{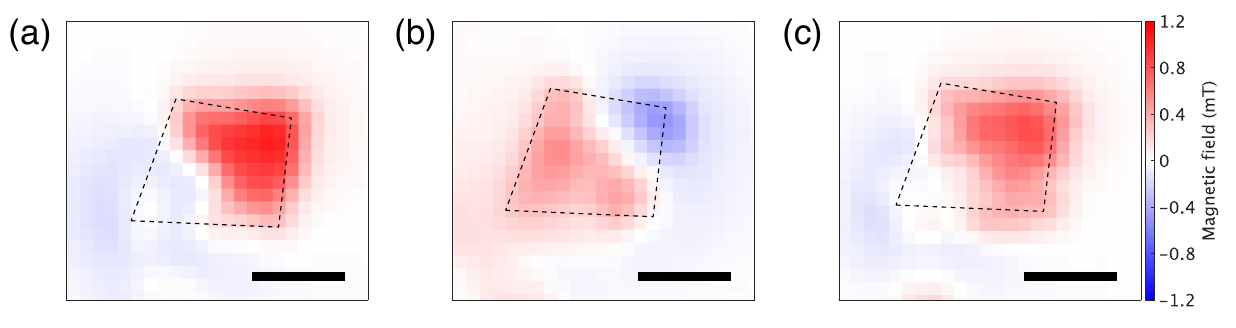}
    \caption{Magnetization reversal of the iron sample. The bias magnetic field has been subtracted. The dashed outline indicates the approximate position of the iron sample. The scale bar corresponds to 20\,$\mathrm{\mu m}$. (a) Stray magnetic field distribution in the initial state. The sample is magnetized to the right by applying an external magnetic field. (b) Stray magnetic field distribution after magnetizing the sample to the left with an external field. The distribution is reversed compared with (a). (c) Stray magnetic field distribution after magnetizing the sample to the right again. The distribution is similar to that in (a).}
    \label{fig3}
\end{figure}

We initially applied an external field in the rightward direction. The resulting stray magnetic field distribution along the $z$-axis, after subtracting the bias field, is shown in Fig.~\ref{fig3}(a). The right side of the iron sample exhibits a stray field pointing out of the page, while the left side exhibits a field pointing into the page, indicating that the magnetization is oriented to the right. Next, an external field was applied in the leftward direction. The corresponding stray-field distribution is shown in Fig.~\ref{fig3}(b), where the right side of the sample shows a field pointing into the page and the left side a field pointing out of the page, demonstrating leftward magnetization. Thus, magnetization reversal in response to the direction of the external field was confirmed.

Finally, the external field was applied again in the rightward direction, and the resulting stray-field distribution is shown in Fig.~\ref{fig3}(c), confirming that the magnetization returned to the rightward direction. These observations verify that the measured magnetic field patterns arise from the stray field of the iron sample rather than from experimental artifacts.

\subsection{Stray field distribution of ferromagnetic-paramagnetic transition}
\begin{figure}
    \centering
    \includegraphics[width=\linewidth]{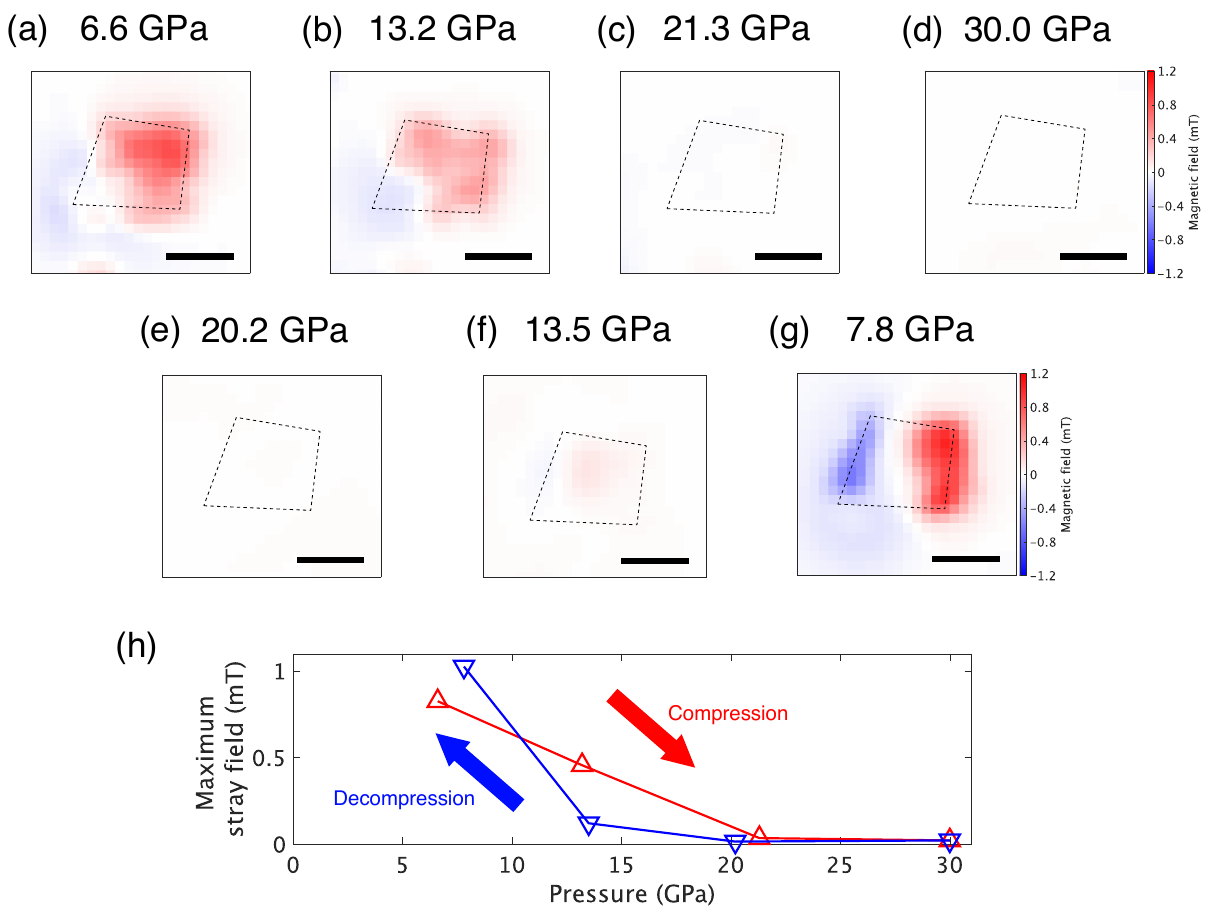}
    \caption{(a–g) Stray magnetic field distributions measured at different pressures. The bias magnetic field has been subtracted. The dashed outline indicates the approximate position of the iron sample. The scale bar corresponds to 20\,$\mathrm{\mu m}$. The disappearance of the stray magnetic field upon compression and its recovery upon decompression are clearly observed. (h) Hysteresis of the stray magnetic field from iron. The horizontal axis represents pressure, and the vertical axis shows the maximum value of the stray magnetic field. The compression and decompression processes are indicated by the red and blue curves, respectively.}
    \label{fig4}
\end{figure}

In order to observe the ferromagnetic–paramagnetic transition of iron, the pressure was increased stepwise to 6.6, 13.2, 21.3, and 30.0\,GPa, and subsequently decreased to 20.2, 13.5, and 7.8\,GPa. At each pressure, the ZFS of the [111]-oriented NV centers differs from that of the other three NV orientations (see Eq.~(\ref{Eq_D})), and the bias magnetic field must therefore be adjusted to prevent overlap of the corresponding ODMR dips.

Ideally, the bias magnetic field should be kept identical for all magnetic measurements. However, because the easy axis of the iron sample lies along the in-plane long axis, magnetization reversal requires magnetic fields on the order of several tens of millitesla, and the applied bias field is oriented normal to the sample plane with a magnitude of only a few millitesla, we conclude that physically meaningful comparisons of the stray magnetic field distributions can be made even when the bias field is adjusted at each pressure.

We first discuss the compression process. The stray magnetic field distribution at 6.6\,GPa was already shown in Fig.~\ref{fig3}(c) and is reproduced here for comparison in Fig.~\ref{fig4}(a). Figure~\ref{fig4}(b) shows the stray magnetic field distribution at 13.2\,GPa (bias field: 1.7\,mT). In this map, the right side of the iron sample exhibits a field pointing out of the page, while the left side exhibits a field pointing into the page, clearly indicating that the iron sample retains a substantial rightward magnetization. Still, the magnitude of the stray magnetic field is reduced compared with that at 6.6\,GPa. Next, Fig.~\ref{fig4}(c) shows the stray magnetic field distribution at 21.3\,GPa (bias field: 2.5\,mT). At this pressure, the stray magnetic field is largely suppressed, indicating that most of the iron sample has transitioned to the paramagnetic state. Nevertheless, a faint residual stray-field pattern remains, with a red region on the right side and a blue region on the left side, suggesting a small remnant rightward magnetization. Finally, Fig.~\ref{fig4}(d) presents the stray magnetic field distribution at 30.0\,GPa (bias field: 1.0\,mT). Here, the stray magnetic field has completely vanished, confirming that the iron sample has fully transitioned to the paramagnetic state. A weak magnetic field distribution is still visible within the field of view; however, this originates from inhomogeneity in the applied bias magnetic field.

We next discuss the decompression process. Figure~\ref{fig4}(e) shows the stray magnetic field distribution at 20.2\,GPa during decompression (bias field: 0.5\,mT). At this pressure, a weak magnetic field distribution is visible; however, as discussed above, this originates from inhomogeneity in the bias magnetic field, and no stray field from the iron sample is observed, indicating that the sample remains in the paramagnetic state. Considering that a small ferromagnetic component was still observed at 21.3\,GPa during compression, this result indicates that the transition back to the ferromagnetic state does not yet occur at pressures lower than the pressure at which the sample becomes fully paramagnetic during compression. This behavior reflects the hysteresis of the $\alpha-\varepsilon$ transition \cite{Lesik:NV_high_pressure,Hsieh:NV_high_pressure}. Upon further decompression, the stray magnetic field distribution at 13.5\,GPa (bias field: 0.6\,mT) is shown in Fig.~\ref{fig4}(f). At this pressure, the recovery of the stray magnetic field is clearly observed. Figure~\ref{fig4}(g) shows the stray magnetic field distribution at 7.8\,GPa (bias field 1.0\,mT). Upon decompression to 7.8\,GPa, the magnitude of the stray magnetic field recovers to a level comparable to that observed at 6.6\,GPa prior to compression, indicating that the iron sample has returned to the ferromagnetic state.

As shown in Fig.~\ref{fig4}(a-g), the use of the (111)-oriented diamond in the present study enables clear stray-field imaging over the entire field of view, including directly above the iron sample, up to 30\,GPa. While previous studies using (100)-oriented anvils demonstrated high-pressure magnetic imaging, a reduction in ODMR contrast at elevated pressures was reported. The present configuration maintains sufficient contrast within this pressure range, allowing reliable stray-field imaging with preserved ODMR contrast.

A comparison between the stray magnetic field distribution at 7.8\,GPa after decompression (Fig.~\ref{fig4}(g)) and that at 6.6\,GPa before compression (Fig.~\ref{fig4}(a)) reveals differences in the spatial distribution. This difference is likely attributable to a change in the shape of the iron sample induced by the compression–decompression cycle.

To examine the pressure dependence of the stray magnetic field, the maximum value of the stray field was plotted as a function of pressure, as shown in Fig.~\ref{fig4}(h). The compression and decompression processes are indicated by the red and blue curves, respectively. As shown in Fig.~\ref{fig4}(h), the pressure dependence of the stray magnetic field exhibits clear hysteresis \cite{Lesik:NV_high_pressure,Hsieh:NV_high_pressure}. This behavior demonstrates that the $\alpha-\varepsilon$ transition of iron is reliably captured in the present measurements.

\section{Conclusion}
In this study, we measured the stray magnetic field distribution of iron under pressures up to 30\,GPa using a (111)-oriented diamond anvil. Upon compression, the stray magnetic field gradually diminished and eventually vanished completely. Upon decompression, the stray magnetic field recovered, allowing us to directly observe the transition of iron from the ferromagnetic to the paramagnetic state and back from the paramagnetic to the ferromagnetic state.

Although NV centers fabricated on the anvil surface enabled measurements of the stray magnetic field from the iron sample under high pressure, NV centers pre-existing inside the diamond anvil were found to act as a background signal. Addressing this issue by establishing a method to eliminate or deactivate internal NV centers is therefore an important challenge for future studies.

At pressures up to approximately 30\,GPa, ODMR signals originating not only from the [111]-oriented NV centers but also from the other three NV orientations are observed, which complicates the spectral analysis. In particular, because the ZFS of the [111]-oriented NV centers shifts due to stress along the [111] direction, it was necessary to carefully adjust the bias magnetic field to avoid overlap between the [111]-derived signal and those from the other three orientations. In addition, the emergence of a pressure gradient across the culet surface upon compression further complicated these adjustments. In contrast, at pressures exceeding approximately 50\,GPa, signals from NV centers other than the [111] orientation are expected to vanish, eliminating the need for such tuning.

By extending this study, which demonstrates stable and quantitative stray-field imaging using (111)-oriented diamond anvils, even more precise magnetic measurements at pressures exceeding 100\,GPa are expected to become feasible. Establishing this robust high-pressure magnetic measurement platform opens the possibility of extending NV-based magnetic imaging to ultrahigh pressures, including the detection of the Meissner effect in hydride high-temperature superconductors\cite{Bhattacharyya:CeH9}.

\section*{Acknowledgements}
This work was partially supported by
JST, CREST Grant No.~JPMJCR23I2, Japan; 
Grants-in-Aid for Scientific Research (Nos.~JP25H01248, JP22K03524, JP23K25800, JP24K21194, JP25K00934, JP25KJ1141, and JP25KJ1166); 
Seiko Instruments Advanced Technology Foundation Research Grants;
Daikin Industry Ltd; 
the Cooperative Research Project of RIEC, Tohoku University;
``Advanced Research Infrastructure for Materials and Nanotechnology in Japan (ARIM)'' (No.~JPMXP1222UT1131) of the Ministry of Education, Culture, Sports, Science and Technology of Japan (MEXT).  
R.S. acknowledge supports from FoPM, WINGS Program, The University of Tokyo.
K.Y. acknowledges supports from JST SPRING Grant No. JPMJSP2108, Japan, and MERIT-WINGS, The University of Tokyo.

\section*{Declaration of interest}
The authors declare no conflict of interest.

\bibliographystyle{ieeetr}
\bibliography{reference}

@article{Lesik:NV_high_pressure,
    author = {Margarita Lesik  and Thomas Plisson  and Loïc Toraille  and Justine Renaud  and Florent Occelli  and Martin Schmidt  and Olivier Salord  and Anne Delobbe  and Thierry Debuisschert  and Loïc Rondin  and Paul Loubeyre  and Jean-François Roch },
    title = {Magnetic measurements on micrometer-sized samples under high pressure using designed NV centers},
    journal = {Science},
    volume = {366},
    number = {6471},
    pages = {1359-1362},
    year = {2019},
    doi = {10.1126/science.aaw4329},
    URL = {https://www.science.org/doi/abs/10.1126/science.aaw4329},
    eprint = {https://www.science.org/doi/pdf/10.1126/science.aaw4329}
}

@article{Hsieh:NV_high_pressure,
    author = {S. Hsieh  and P. Bhattacharyya  and C. Zu  and T. Mittiga  and T. J. Smart  and F. Machado  and B. Kobrin  and T. O. Höhn  and N. Z. Rui  and M. Kamrani  and S. Chatterjee  and S. Choi  and M. Zaletel  and V. V. Struzhkin  and J. E. Moore  and V. I. Levitas  and R. Jeanloz  and N. Y. Yao },
    title = {Imaging stress and magnetism at high pressures using a nanoscale quantum sensor},
    journal = {Science},
    volume = {366},
    number = {6471},
    pages = {1349-1354},
    year = {2019},
    doi = {10.1126/science.aaw4352},
    URL = {https://www.science.org/doi/abs/10.1126/science.aaw4352},
    eprint = {https://www.science.org/doi/pdf/10.1126/science.aaw4352}
}

@article{Doherty:pressure,
  title = {Electronic Properties and Metrology Applications of the Diamond ${\mathrm{NV}}^{\ensuremath{-}}$ Center under Pressure},
  author = {Doherty, Marcus W. and Struzhkin, Viktor V. and Simpson, David A. and McGuinness, Liam P. and Meng, Yufei and Stacey, Alastair and Karle, Timothy J. and Hemley, Russell J. and Manson, Neil B. and Hollenberg, Lloyd C. L. and Prawer, Steven},
  journal = {Phys. Rev. Lett.},
  volume = {112},
  issue = {4},
  pages = {047601},
  numpages = {5},
  year = {2014},
  month = {Jan},
  publisher = {American Physical Society},
  doi = {10.1103/PhysRevLett.112.047601},
  url = {https://link.aps.org/doi/10.1103/PhysRevLett.112.047601}
}

@article{Hilberer:Pressure,
  title = {Enabling quantum sensing under extreme pressure: Nitrogen-vacancy magnetometry up to 130 {GPa}},
  author = {Hilberer, Antoine and Toraille, Lo\"{\i}c and Dailledouze, Cassandra and Adam, Marie-Pierre and Hanlon, Liam and Weck, Gunnar and Schmidt, Martin and Loubeyre, Paul and Roch, Jean-Fran\ifmmode \mbox{\c{c}}\else \c{c}\fi{}ois},
  journal = {Phys. Rev. B},
  volume = {107},
  issue = {22},
  pages = {L220102},
  numpages = {6},
  year = {2023},
  month = {Jun},
  publisher = {American Physical Society},
  doi = {10.1103/PhysRevB.107.L220102},
  url = {https://link.aps.org/doi/10.1103/PhysRevB.107.L220102}
}

@Article{Bhattacharyya:CeH9,
author={Bhattacharyya, P.
and Chen, W.
and Huang, X.
and Chatterjee, S.
and Huang, B.
and Kobrin, B.
and Lyu, Y.
and Smart, T. J.
and Block, M.
and Wang, E.
and Wang, Z.
and Wu, W.
and Hsieh, S.
and Ma, H.
and Mandyam, S.
and Chen, B.
and Davis, E.
and Geballe, Z. M.
and Zu, C.
and Struzhkin, V.
and Jeanloz, R.
and Moore, J. E.
and Cui, T.
and Galli, G.
and Halperin, B. I.
and Laumann, C. R.
and Yao, N. Y.},
title={Imaging the Meissner effect in hydride superconductors using quantum sensors},
journal={Nature},
year={2024},
month={Mar},
day={01},
volume={627},
number={8002},
pages={73-79},
issn={1476-4687},
doi={10.1038/s41586-024-07026-7},
url={https://doi.org/10.1038/s41586-024-07026-7}
}

@article{Goldman:Contrast_theory,
  title = {State-selective intersystem crossing in nitrogen-vacancy centers},
  author = {Goldman, M. L. and Doherty, M. W. and Sipahigil, A. and Yao, N. Y. and Bennett, S. D. and Manson, N. B. and Kubanek, A. and Lukin, M. D.},
  journal = {Phys. Rev. B},
  volume = {91},
  issue = {16},
  pages = {165201},
  numpages = {11},
  year = {2015},
  month = {Apr},
  publisher = {American Physical Society},
  doi = {10.1103/PhysRevB.91.165201},
  url = {https://link.aps.org/doi/10.1103/PhysRevB.91.165201}
}

@article{Yagi:DAC,
author = {Takehiko Yagi and Takeshi Sakai and Hirokazu Kadobayashi and Tetsuo Irifune},
title = {Review: high pressure generation techniques beyond the limit of conventional diamond anvils},
journal = {High Pressure Research},
volume = {40},
number = {1},
pages = {148--161},
year = {2020},
publisher = {Taylor \& Francis},
doi = {10.1080/08957959.2019.1704753},
URL = {https://doi.org/10.1080/08957959.2019.1704753},
eprint = {https://doi.org/10.1080/08957959.2019.1704753},
}

@article{Mao:Ruby,
    author = {Mao, H. K. and Bell, P. M. and Shaner, J. W. and Steinberg, D. J.},
    title = {Specific volume measurements of {Cu}, {Mo}, {Pd}, and {Ag} and calibration of the ruby {R1} fluorescence pressure gauge from 0.06 to 1 {M}bar},
    journal = {Journal of Applied Physics},
    volume = {49},
    number = {6},
    pages = {3276-3283},
    year = {1978},
    month = {06},
    issn = {0021-8979},
    doi = {10.1063/1.325277},
    url = {https://doi.org/10.1063/1.325277},
    eprint = {https://pubs.aip.org/aip/jap/article-pdf/49/6/3276/18380627/3276_1_online.pdf},
}

@Article{Drozdov:H3S,
author={Drozdov, A. P.
and Eremets, M. I.
and Troyan, I. A.
and Ksenofontov, V.
and Shylin, S. I.},
title={Conventional superconductivity at 203 kelvin at high pressures in the sulfur hydride system},
journal={Nature},
year={2015},
month={Sep},
day={01},
volume={525},
number={7567},
pages={73-76},
doi={10.1038/nature14964},
url={https://doi.org/10.1038/nature14964}
}

@Article{Einaga:H3S_structure,
author={Einaga, Mari
and Sakata, Masafumi
and Ishikawa, Takahiro
and Shimizu, Katsuya
and Eremets, Mikhail I.
and Drozdov, Alexander P.
and Troyan, Ivan A.
and Hirao, Naohisa
and Ohishi, Yasuo},
title={Crystal structure of the superconducting phase of sulfur hydride},
journal={Nature Physics},
year={2016},
month={Sep},
day={01},
volume={12},
number={9},
pages={835-838},
issn={1745-2481},
doi={10.1038/nphys3760},
url={https://doi.org/10.1038/nphys3760}
}

@Article{Drozdov:LaH10,
author={Drozdov, A. P.
and Kong, P. P.
and Minkov, V. S.
and Besedin, S. P.
and Kuzovnikov, M. A.
and Mozaffari, S.
and Balicas, L.
and Balakirev, F. F.
and Graf, D. E.
and Prakapenka, V. B.
and Greenberg, E.
and Knyazev, D. A.
and Tkacz, M.
and Eremets, M. I.},
title={Superconductivity at 250 {K} in lanthanum hydride under high pressures},
journal={Nature},
year={2019},
month={May},
day={01},
volume={569},
number={7757},
pages={528-531},
issn={1476-4687},
doi={10.1038/s41586-019-1201-8},
url={https://doi.org/10.1038/s41586-019-1201-8}
}

@article{Somayazalu:LaH10,
  title = {Evidence for Superconductivity above 260 {K} in Lanthanum Superhydride at Megabar Pressures},
  author = {Somayazulu, Maddury and Ahart, Muhtar and Mishra, Ajay K. and Geballe, Zachary M. and Baldini, Maria and Meng, Yue and Struzhkin, Viktor V. and Hemley, Russell J.},
  journal = {Phys. Rev. Lett.},
  volume = {122},
  issue = {2},
  pages = {027001},
  numpages = {6},
  year = {2019},
  month = {Jan},
  publisher = {American Physical Society},
  doi = {10.1103/PhysRevLett.122.027001},
  url = {https://link.aps.org/doi/10.1103/PhysRevLett.122.027001}
}

@article{Taylor:Fe_Hysteresis,
    author = {Taylor, R. D. and Pasternak, M. P. and Jeanloz, R.},
    title = {Hysteresis in the high pressure transformation of bcc‐ to hcp‐iron},
    journal = {Journal of Applied Physics},
    volume = {69},
    number = {8},
    pages = {6126-6128},
    year = {1991},
    month = {04},
    issn = {0021-8979},
    doi = {10.1063/1.348779},
    url = {https://doi.org/10.1063/1.348779},
    eprint = {https://pubs.aip.org/aip/jap/article-pdf/69/8/6126/18642433/6126_1_online.pdf},
}

@Article{Eremets:review,
author={Eremets, M. I.
and Minkov, V. S.
and Drozdov, A. P.
and Kong, P. P.
and Ksenofontov, V.
and Shylin, S. I.
and Bud'ko, S. L.
and Prozorov, R.
and Balakirev, F. F.
and Sun, Dan
and Mozaffari, S.
and Balicas, L.},
title={High-Temperature Superconductivity in Hydrides: Experimental Evidence and Details},
journal={Journal of Superconductivity and Novel Magnetism},
year={2022},
month={Apr},
day={01},
volume={35},
number={4},
pages={965-977},
issn={1557-1947},
doi={10.1007/s10948-022-06148-1},
url={https://doi.org/10.1007/s10948-022-06148-1}
}

@Article{Taylor2008:NV,
author={Taylor, J. M.
and Cappellaro, P.
and Childress, L.
and Jiang, L.
and Budker, D.
and Hemmer, P. R.
and Yacoby, A.
and Walsworth, R.
and Lukin, M. D.},
title={High-sensitivity diamond magnetometer with nanoscale resolution},
journal={Nature Physics},
year={2008},
month={Oct},
day={01},
volume={4},
number={10},
pages={810-816},
issn={1745-2481},
doi={10.1038/nphys1075},
url={https://doi.org/10.1038/nphys1075}
}

@article{Degen2008:NV,
    author = {Degen, C. L.},
    title = {Scanning magnetic field microscope with a diamond single-spin sensor},
    journal = {Applied Physics Letters},
    volume = {92},
    number = {24},
    pages = {243111},
    year = {2008},
    month = {06},
    issn = {0003-6951},
    doi = {10.1063/1.2943282},
    url = {https://doi.org/10.1063/1.2943282},
    eprint = {https://pubs.aip.org/aip/apl/article-pdf/doi/10.1063/1.2943282/14395578/243111_1_online.pdf},
}

@Article{Maze2008:NV,
author={Maze, J. R.
and Stanwix, P. L.
and Hodges, J. S.
and Hong, S.
and Taylor, J. M.
and Cappellaro, P.
and Jiang, L.
and Dutt, M. V. Gurudev
and Togan, E.
and Zibrov, A. S.
and Yacoby, A.
and Walsworth, R. L.
and Lukin, M. D.},
title={Nanoscale magnetic sensing with an individual electronic spin in diamond},
journal={Nature},
year={2008},
month={Oct},
day={01},
volume={455},
number={7213},
pages={644-647},
issn={1476-4687},
doi={10.1038/nature07279},
url={https://doi.org/10.1038/nature07279}
}

@Article{Balasubramanian2008:NV,
author={Balasubramanian, Gopalakrishnan
and Chan, I. Y.
and Kolesov, Roman
and Al-Hmoud, Mohannad
and Tisler, Julia
and Shin, Chang
and Kim, Changdong
and Wojcik, Aleksander
and Hemmer, Philip R.
and Krueger, Anke
and Hanke, Tobias
and Leitenstorfer, Alfred
and Bratschitsch, Rudolf
and Jelezko, Fedor
and Wrachtrup, J{\"o}rg},
title={Nanoscale imaging magnetometry with diamond spins under ambient conditions},
journal={Nature},
year={2008},
month={Oct},
day={01},
volume={455},
number={7213},
pages={648-651},
issn={1476-4687},
doi={10.1038/nature07278},
url={https://doi.org/10.1038/nature07278}
}

@article{Acosta:NV_temperature,
  title = {Temperature Dependence of the Nitrogen-Vacancy Magnetic Resonance in Diamond},
  author = {Acosta, V. M. and Bauch, E. and Ledbetter, M. P. and Waxman, A. and Bouchard, L.-S. and Budker, D.},
  journal = {Phys. Rev. Lett.},
  volume = {104},
  issue = {7},
  pages = {070801},
  numpages = {4},
  year = {2010},
  month = {Feb},
  publisher = {American Physical Society},
  doi = {10.1103/PhysRevLett.104.070801},
  url = {https://link.aps.org/doi/10.1103/PhysRevLett.104.070801}
}

@Article{Kucsko:NV_temperature,
author={Kucsko, G.
and Maurer, P. C.
and Yao, N. Y.
and Kubo, M.
and Noh, H. J.
and Lo, P. K.
and Park, H.
and Lukin, M. D.},
title={Nanometre-scale thermometry in a living cell},
journal={Nature},
year={2013},
month={Aug},
day={01},
volume={500},
number={7460},
pages={54-58},
issn={1476-4687},
doi={10.1038/nature12373},
url={https://doi.org/10.1038/nature12373}
}

@article{Yamamoto:NV_temperature,
doi = {10.35848/1882-0786/adac2a},
url = {https://doi.org/10.35848/1882-0786/adac2a},
year = {2025},
month = {feb},
publisher = {IOP Publishing},
volume = {18},
number = {2},
pages = {025001},
author = {Yamamoto, Kouki and Ogawa, Kensuke and Tsukamoto, Moeta and Ashida, Yuto and Sasaki, Kento and Kobayashi, Kensuke},
title = {Nanodiamond quantum thermometry assisted with machine learning},
journal = {Applied Physics Express}
}

@article{Tetienne:NV_DW,
author = {J.-P. Tetienne  and T. Hingant  and J.-V. Kim  and L. Herrera Diez  and J.-P. Adam  and K. Garcia  and J.-F. Roch  and S. Rohart  and A. Thiaville  and D. Ravelosona  and V. Jacques },
title = {Nanoscale imaging and control of domain-wall hopping with a nitrogen-vacancy center microscope},
journal = {Science},
volume = {344},
number = {6190},
pages = {1366-1369},
year = {2014},
doi = {10.1126/science.1250113},
URL = {https://www.science.org/doi/abs/10.1126/science.1250113},
eprint = {https://www.science.org/doi/pdf/10.1126/science.1250113}
}

@Article{Velez:NV_DW,
author={V{\'e}lez, Sa{\"u}l
and Schaab, Jakob
and W{\"o}rnle, Martin S.
and M{\"u}ller, Marvin
and Gradauskaite, Elzbieta
and Welter, Pol
and Gutgsell, Cameron
and Nistor, Corneliu
and Degen, Christian L.
and Trassin, Morgan
and Fiebig, Manfred
and Gambardella, Pietro},
title={High-speed domain wall racetracks in a magnetic insulator},
journal={Nature Communications},
year={2019},
month={Oct},
day={18},
volume={10},
number={1},
pages={4750},
issn={2041-1723},
doi={10.1038/s41467-019-12676-7},
url={https://doi.org/10.1038/s41467-019-12676-7}
}

@Article{Tsukamoto:NV_ML_mag,
author={Tsukamoto, Moeta
and Ito, Shuji
and Ogawa, Kensuke
and Ashida, Yuto
and Sasaki, Kento
and Kobayashi, Kensuke},
title={Accurate magnetic field imaging using nanodiamond quantum sensors enhanced by machine learning},
journal={Scientific Reports},
year={2022},
month={Sep},
day={01},
volume={12},
number={1},
pages={13942},
issn={2045-2322},
doi={10.1038/s41598-022-18115-w},
url={https://doi.org/10.1038/s41598-022-18115-w}
}

@article{Suda:ND_Pressure,
author = {Suda ,Ryotaro and Uriu ,Kenshin and Yamamoto ,Kouki and Sasaki ,Misaki and Sasaki ,Kento and Einaga ,Mari and Shimizu ,Katsuya and Kobayashi ,Kensuke},
title = {{GPa} Pressure Imaging Using Nanodiamond Quantum Sensors},
journal = {Journal of the Physical Society of Japan},
volume = {94},
number = {12},
pages = {124707},
year = {2025},
doi = {10.7566/JPSJ.94.124707},
URL = {https://doi.org/10.7566/JPSJ.94.124707},
eprint = {https://doi.org/10.7566/JPSJ.94.124707}
}

@article{Mai:NV_111_iron,
    author = {Mai, Di and Zhong, Cheng and Wang, Ziqi and Wang, He and Sun, Xiaoyu and Dai, Rucheng and Wang, Zhongping and Zhang, Zengming},
    title = {Megabar pressure sensing and magnetic phase imaging by [111]-oriented nitrogen-vacancy centers in diamond},
    journal = {Journal of Applied Physics},
    volume = {138},
    number = {4},
    pages = {045901},
    year = {2025},
    month = {07},
    issn = {0021-8979},
    doi = {10.1063/5.0278258},
    url = {https://doi.org/10.1063/5.0278258},
    eprint = {https://pubs.aip.org/aip/jap/article-pdf/doi/10.1063/5.0278258/20605425/045901_1_5.0278258.pdf},
}

@Article{Wang:NV_111_magnetite,
author={Wang, Mengqi
and Wang, Yu
and Liu, Zhixian
and Xu, Ganyu
and Yang, Bo
and Yu, Pei
and Sun, Haoyu
and Ye, Xiangyu
and Zhou, Jingwei
and Goncharov, Alexander F.
and Wang, Ya
and Du, Jiangfeng},
title={Imaging magnetic transition of magnetite to megabar pressures using quantum sensors in diamond anvil cell},
journal={Nature Communications},
year={2024},
month={Oct},
day={14},
volume={15},
number={1},
pages={8843},
issn={2041-1723},
doi={10.1038/s41467-024-52272-y},
url={https://doi.org/10.1038/s41467-024-52272-y}
}

\end{document}